\documentclass[fleqn,usenatbib]{mnras}

\usepackage{mathptmx}
\usepackage{float}

\usepackage[T1]{fontenc}
\usepackage{ae,aecompl}

\usepackage{graphicx}	% Including figure files
\usepackage{amsmath}	% Advanced maths commands
\usepackage{amssymb}
\usepackage{amsfonts}% Extra maths symbols
\usepackage[normalem]{ulem}
\usepackage{soul}
\usepackage{multirow}
\usepackage{mathtools}
\usepackage{mismath}
\usepackage{lmodern}
\usepackage{xspace} % this decides if space is needed or not

\newcommand{\sumnn}{\sum_{j\in \text{NN}(i)}}

\newcommand{\PG}{{\sc p-gadget}3\xspace}
\newcommand{\AG}{{\sc ax-gadget}\xspace}
\newcommand{\FGAS}{{\sc fuzzy-gasoline}\xspace}

\newcommand{\ETC}{{\sc gasoline2}\xspace}

\newcommand{\AC}{{\sc axionCAMB}\xspace}

\renewcommand{\vec}[1]{\mathbf{#1}}

\title[Fuzzy Galaxies]{Fuzzy Gasoline: Cosmological hydrodynamical simulations of dwarf galaxy formation with Fuzzy Dark Matter}

\author[M. Nori et al.]{
Matteo Nori,$^{1,2}$\thanks{E-mail: matteo.nori@nyu.edu}, Shubhan Bhatia$^{1,2}$ and Andrea V. Macciò$^{1,2,3}$
\\
$^{1}$New York University Abu Dhabi, PO Box 129188 Saadiyat Island, Abu Dhabi, United Arab Emirates\\
$^{2}$Center for Astrophysics and Space Science (CASS), New York University Abu Dhabi, United Arab Emirates\\
$^{3}$Max Planck Institut f\"{u}r Astronomie, K\"{o}nigstuhl 17, D-69117 Heidelberg, Germany\\
}

\date{Accepted XXX. Received YYY; in original form ZZZ}

\pubyear{2024}
\begin{document}
\label{firstpage}
\pagerange{\pageref{firstpage}--\pageref{lastpage}}
\maketitle

\begin{abstract}
\setlength{\parindent}{0pt} % Remove indent for the abstract
\newline
We present the first set of high-resolution, hydrodynamical cosmological simulations of galaxy formation in a Fuzzy Dark Matter (FDM) framework. These simulations were performed with a new version of the {\sc gasoline2} code, known as \FGAS, which can simulate quantum FDM effects alongside a comprehensive baryonic model that includes metal cooling, star formation, supernova feedback, and black hole physics, previously used in the NIHAO simulation suite.
Using thirty zoom-in simulations of galaxies with halo masses in the range $10^9 \lesssim M_{\text{halo}}/M_{\odot} \lesssim 10^{11}$, we explore how the interplay between FDM’s quantum potential and baryonic processes influences dark matter distributions and observable galaxy properties. Our findings indicate that both baryons and low-mass FDM contribute to core formation within dark matter profiles, though through distinct mechanisms: FDM-induced cores emerge in all haloes, particularly within low-mass systems at high redshift, while baryon-driven cores form within a specific mass range and at low redshift.
Despite these significant differences in dark matter structure, key stellar observables such as star formation histories and velocity dispersion profiles remain remarkably similar to predictions from the Cold Dark Matter (CDM) model, making it challenging to distinguish between CDM and FDM solely through stellar observations.

\end{abstract}

\begin{keywords}
cosmology: theory -- methods: numerical
\end{keywords}

\section{Introduction}
\label{sec:intro}

The Cold Dark Matter (CDM) model, characterized by its \textit{cold}, \textit{dark}, and \textit{collisionless} nature, has been considered the leading framework for explaining the dark matter component in cosmic structure formation over the past few decades \citep[see e.g.][for a comprehensive review on the subject]{2010gfe..book.....M}. Nonetheless, unresolved tensions at small scales, combined with the ongoing failure to detect Weakly Interacting Massive Particles (WIMPs) — the leading particle candidate of the CDM model — have continued to raise doubts about the model’s viability. Motivated by the elusiveness of WIMPs in predominant direct and indirect detection methods, several alternative dark matter models have come to the forefront, investigating the lower mass regimes for dark matter particles \citep{Jungman95}. Moving away from the GeV/c$^2$ mass range associated with WIMPs, these efforts explored and proposed several lighter dark matter particle candidates, one being the axion particle, which is theorized to arise from the CP-symmetry breaking in quantum chromo-dynamics (QCD) theories \citep{Peccei_Quinn_1977}.

In a cosmological context, a pseudo-scalar bosonic particle can be generalized from the QCD axion model, motivating a comprehensive class of axion-like particles (ALPs) acting as dark matter candidates. These ALPs span a broad range of masses, encompassing over 24 orders of magnitude from \(10^{-24}\) to 1 eV/c$^2$ \citep[see e.g.][for reviews on FDM models]{2017PhRvD..95d3541H,2021A&ARv..29....7F}. Dark matter models related to ALP particle masses in the mass range (\(10^{-24}\) to \(10^{-19}\) eV/c$^2$) are known as \textit{Fuzzy Dark Matter} (FDM) models, whose identifying boson mass $m_\chi$ is typically represented in terms of $m_{22} = m_{\chi} / (10^{-22} \text{ eV}/c^2)$. The mass range of FDM corresponds to de Broglie wavelengths on scales of \(\mathcal{O}(1 \text{ kpc})\), exhibiting wave-like behavior at sub-galactic scales \citep{2000PhRvL..85.1158H}.

The quantum wave-like nature of FDM results in a net repulsive force that, on one hand, modifies the matter power spectrum of cold dark matter (CDM) during matter-radiation equality and smooths out density perturbations at small scales, ultimately leading to fewer collapsed structures \citep{2000PhRvL..85.1158H, 2014MNRAS.437.2652M}. On the other hand, it induces a resistance to gravitational collapse resulting in decreased dark matter (DM) distribution in the central region of FDM haloes. This effectively translates to FDM haloes featuring cored inner DM density profiles ($\rho(r) \sim$ constant) contrasted with CDM's cuspy inner DM density profiles ($\rho(r) \sim r^{-1}$) for dwarf galaxy systems \citep{2000PhRvL..85.1158H}. 

While the CDM model has been successful in modeling large-scale cosmological structures \citep{2005Natur.435..629S, PhysRevD.74.123507, 2017MNRAS.470.2617A}, several challenges have arisen on smaller, non-linear scales. These include well-known issues such as the cusp-core problem \citep{flores1994observational, Moore_1994} and the missing satellites problem \citep{1999ApJ...522...82K, Moore_1999} [see \cite{2017ARA&A..55..343B} for a detailed review]. Verifying the model’s validity at these non-linear scales has proven to be particularly challenging. In response, numerous studies have defended the CDM model, pointing out that earlier works overstated the severity of these problems due to theoretical and observational limitations. These studies emphasize the growing importance of baryonic physics in structure formation on smaller scales \citep{2014ApJ...786...87B, 2020MNRAS.495L..46M, 2022MNRAS.514.5307W}, as well as the inefficiency of star formation in dwarf galaxies, which complicates their observational detection \citep{2003MNRAS.339.1057Y,2017MNRAS.471.3547F,Frings2017MNRAS}. Previous studies investigating the role of baryonic feedback processes in dwarf galaxies have found that baryons are able to produce significant cores ($\sim$ 1 kpc) in their dark matter distribution \citep{2010Natur.463..203G, Macciò_2012, 2019MNRAS.488.2387B}. The most-widely accepted mechanism explaining this phenomenon is the sub-dynamical time-scale changes in the central ($\sim \bigO(\text{kpc})$) potential of the halo. These rapid changes in the central potential, caused by stellar and black hole feedback, are tied to strong gas outflows that irreversibly alter the central potential by transferring energy to collisionless DM particles \citep{2012MNRAS.421.3464P}. However, these baryonic effects help alleviate these small scale tensions only up to a certain mass scale ($M_{halo}$ $\sim$ 10$^{10} M_{\odot}$). Since these mechanisms are out of play in lowest mass, gas-deficient dark-matter dominated dwarf galaxies, the central DM distribution of the halo reverts back to the cuspy profiles \citep[e.g.][]{Tollet15}. The addition of FDM interaction to baryonic effects might help alleviate these tensions at lower halo masses while maintaining CDM large scale features.

Numerical simulations of structure formation within FDM models have been initially performed by means of highly numerically intensive Adaptive Mesh Refinement (AMR) algorithms able to solve the Schr\"odinger-Poisson equations over a grid \citep[see e.g.][]{GAMER,GAMER2,Mocz17}, leading to impressive and very detailed results on the properties of individual FDM collapsed objects \citep[see e.g.][]{Woo09,Schive14,Veltmaat18}. However, the computational cost of such approach hindered the possibility to extend the investigation of late time structure formation to large cosmological volumes. To address this issue, N-Body codes were employed, initially only including the (linear) suppression in the initial conditions but neglecting the integrated effect of the FDM interaction during the subsequent dynamical evolution \citep[see e.g.][]{Schive16,Irsic17,Armengaud17} --~i.e. basically treating FDM as standard dark matter with a suppressed primordial power spectrum.
The inclusion of the typical FDM interaction in N-body codes was achieved with \AG \citet{Nori18}, a modified version of the cosmological hydrodynamical code \PG that implemented the general scheme suggested by \citet{Mocz15} and \citet{Marsh15}. The code \AG allowed the investigation of FDM in larger cosmological volumes with a vast number of systems \citep{Nori19} as well as a variety of complex galactic systems with many evolving substructures \citep{Nori20,Nori23,elgamal2024no} hardly obtainable with other simulation strategies. Nonetheless, previous studies on FDM cosmologies with \AG have all relied on dark-matter-only (DMO) simulations.

To further investigate FDM models in a proper physical context and examine their impact on galaxy formation, this work expands on what has been done with \AG since \citet{Nori18} by incorporating baryonic effects in a cosmological hydrodynamical N-body code, which are essential for a correct description of structure formation. 

While effective in modeling FDM behavior, \AG is limited in simulating baryonic processes like gas cooling, star formation, and black hole feedback. Conversely, \ETC --~another cosmological hydrodynamical code with a compatible N-body structure~-- has been constantly developed and integrated with new routines related to baryonic process, and has been shown to be very effective in these areas in the past years \citep[e.g.][]{2006MNRAS.373.1074S,Brooks13,wang2015nihao}. By integrating the FDM routines from \AG into \ETC \citep{2017MNRAS.471.2357W}, we have developed a new version of the \ETC code, \FGAS, capable of running hydrodynamic FDM simulations with baryons through to the present day ($z = 0$) of large and complex systems at a reasonable computational cost. To the authors knowledge, this is the first code of its kind capable to do so.

In this work, we leverage the \FGAS code to create novel hydrodynamical simulations of dwarf galaxy systems with halo masses in the range of \(10^9 \lesssim M_{\text{halo}}/M_{\odot} \lesssim 10^{11}\). We detail their properties, including dark matter, gas and star density and velocity profiles, as well as star formation histories, and compare them with those of their cold dark matter (CDM) NIHAO counterparts. Our goal is to explore two key aspects: first, what is the combined effect of baryons and FDM on galactic properties, and second, whether it is possible to disentangle the degeneracy of the two individual contributions.

\quad

The remainder of this paper is organized as follows: in Sec.~\ref{sec:theory} we provide an overview of the theoretical background of FDM models; in Sec.~\ref{sec:NM} we detail the numerical methodology implemented in this work, specifically related to FDM dynamics and simulations; in Sec.~\ref{sec:results} we present the main results, focusing on DM density profiles, differentiating between its two driving factors: FDM's quantum pressure and baryonic feedback processes, and their impact on the observable properties of the explored systems; finally, we summarize our findings in Sec.~\ref{sec:conclusions}.

\section{Theory}
\label{sec:theory}

\subsection{Fuzzy Dark Matter models}
\label{sec:fdm_th}

The intrinsically quantum nature of FDM, representing an ultralight scalar particle model, it is described using a quantum bosonic field \(\hat{\phi}\) \citep{2017PhRvD..95d3541H,2000PhRvL..85.1158H}. The evolution of this bosonic field follows the Gross-Pitaevskii-Poisson equation \citep{1961NCim...20..454G,Pitaevskii61}:
\begin{equation}
    \label{eq:GPP}
    i \frac{\hslash}{m_{\chi}} \frac{\partial \hat{\phi}}{\partial t} = - \frac{\hslash^2}{m_{\chi}^2} \nabla^2 \hat{\phi} + \Phi \hat{\phi}
\end{equation}

where \(m_{\chi}\) is the FDM particle mass, and \(\Phi\) represents the Newtonian gravitational potential. By applying the Madelung transformations \citep{madelung1927quantum}, it is possible to convert this field description into a fluid description, transforming the bosonic field as:
\begin{equation}
    \hat{\phi} = \sqrt{\frac{\rho}{m_{\chi}}} e^{i \frac{\theta}{\hslash}}
\end{equation}

where $\rho$ is the fluid density and $\theta$ is the phase parameter, related to the fluid velocity $\vec{v}$ by $\mathbf{v} = \nabla \theta / m_{\chi}$. The fluid density and velocity can be conversely expressed in terms of field as:
\begin{gather}
\rho = |\hat{\phi}|^2, \\
\mathbf{v} = \frac{\hslash}{m_{\chi}} \Im\left( \frac{\nabla \hat{\phi}}{\hat{\phi}} \right).
\end{gather}

The comoving distance is described using \(\mathbf{x}\) and the comoving velocity is described using \(\mathbf{u}\) which is the comoving equivalent of the fluid velocity \(\mathbf{v}\). The real and imaginary components of equation (\ref{eq:GPP}) transform into the following continuity and modified Euler equations respectively: 
\begin{equation}
\begin{aligned}
\dot{\rho} + 3H\rho + \nabla \cdot (\rho \mathbf{u}) &= 0 \\
\dot{\mathbf{u}} + 2H \mathbf{u} + (\mathbf{u} \cdot \nabla) \mathbf{u} &= - \frac{\nabla \Phi}{a^2} + \frac{\nabla Q}{a^4}
\end{aligned}
\label{eq:MEuler}
\end{equation}

where $\Phi$ is the gravitational potential and satisfies the typical Poisson equation:
\begin{equation}
    \label{eq:Poisson}
    \nabla^2 \Phi = 4 \pi G a^2 \rho_b \delta
\end{equation}

where $\delta = (\rho - \rho_b)/\rho_b$ is the density contrast with respect to the background field density $\rho_b$ \citep{Peebles80}.

The modified Euler equation (Eq.~\ref{eq:MEuler}) features an additional quantum potential component alluded to in the introduction. This Quantum Potential $Q$ (QP, hereafter) is defined as follows:
\begin{equation}
    Q = \frac{\hslash^2}{2m_{\chi}^2} \frac{\nabla^2 \sqrt{\rho}}{\sqrt{\rho}} = \frac{\hslash^2}{2m_{\chi}^2} \left( \frac{\nabla^2 \rho}{2 \rho} - \frac{|\nabla \rho|^2}{4\rho^2} \right)
\end{equation}

and accounts for the quantum nature of the FDM field \citep{Bohm52}. 

The system described by a combination of the Poisson equation and the modified Euler equation of Eq.~\ref{eq:MEuler} can be regarded as a modified Euler-Poisson equation (mEP) system.

One must note that, in principle, the QP should appear in the usual WIMP CDM cosmology in its Euler equation as well, as soon as the assumptions regarding the classical limit is dropped. However, the large particle masses prescribed by the WIMP model makes the $\frac{\hslash^2}{2m_{\chi}^2}$ factor negligible.

Stable solutions of the mEP system have no analytical form but feature a non-divergent central density. The ground state solution is usually referred to as \textit{soliton} \citep{2011PhRvD..84d3531C,2017PhRvD..95d3541H}. The solitonic solution can be approximated as:
\begin{equation}
\rho(r) = \rho_c \left[ 1 + (\sqrt[8]{2} - 1) \left( \frac{r}{r_c} \right)^2 \right]^{-8}
\end{equation}
where the two parameters $\rho_c = \rho(r = 0)$ and $r_c : \rho(r_c) = \frac{\rho_c}{2}$ represent the core density and the core radius, respectively \citep{2000PhRvL..85.1158H}.

On the contrary, it is widely accepted that the density profiles of CDM haloes are well characterized by a central divergence, parametrized in the NFW profile \citep{1996MNRAS.283L..72N}: 
\begin{equation}
    \rho(r) = \rho_s \left( \frac{r}{r_s} \right)^{-1} \left( 1 + \frac{r}{r_s} \right)^{-2}
\end{equation}
where $r_s$ and $\rho_s$ are the characteristic radius and density scale, satisfying the condition $\rho_s = \rho(r=r_s) / 4$.

Since the QP is able to take over the gravitational potential only at small scales, FDM haloes exhibit a hybrid dark matter density profile, which is consistent with an FDM cored solution in the center whereas revert to a usual NFW profiles elsewhere. Thus, the overall profile of FDM haloes can be written as:

\begin{equation}
    \rho(r) = 
    \begin{cases}
        \rho_c \left[ 1 + (\sqrt[8]{2} - 1) \left( \frac{r}{r_c} \right)^2 \right]^{-8} & r < r_t \\
        \rho_s \left( \frac{r}{r_s} \right)^{-1} \left( 1 + \frac{r}{r_s} \right)^{-2} & r \ge r_t
    \end{cases}
\end{equation}
where $r_t$ is the transition radius where the profile shifts from the cored-FDM profile to the typical NFW profile \citep{Schive14,2022MNRAS.511..943C}.

\section{Numerical Methods}
\label{sec:NM}

%In this Section, we introduce and describe the simulations presented in this work and the numerical algorithms used in the simulation process.

\subsection{Implementation of FDM dynamics}

In this Section, we describe the implementation scheme used in \FGAS to represent FDM dynamics, largely based on the one of \AG \citep{Nori18}, focusing on similarities and improvements of the new scheme.

The implementation that calculates the quantities relevant to FDM dynamics revolves around a modified version of the Smoothed Particle Hydrodynamic (SPH) routines. This is a well-known and widespread scheme that is used to infer continuous quantities from a discrete sample, by assigning a virtual volume of representation to each discrete element. The great strength of any SPH scheme relies on its local nature, as calculations at any point only involve a subset of neighboring particles, thus making it extensively used in N-body codes.

Both \PG and \ETC rely on SPH routines to compute hydrodynamic quantities related to gas (and star) particles, although the two codes still retain some differences in the way they allocate particle quantities and iterate over them in the SPH loops. 

Following the initial suggestion of evaluating FDM dynamic quantities via SPH put forward by \citet{Mocz15} and \citet{Marsh15}, it is possible design a modified SPH routine that can keep track of FDM relevant quantities. 

The main FDM specific SPH scheme, shared by both \AG and \FGAS codes, can be briefly described as in the following \citep[refer to][for an in depth description of the scheme]{Nori18}. 

Continuous quantities of interest at the position of particle $i$ are calculated as summations over a given number (chosen as an initial input parameter) of $j$ particles that are \textit{neighbors} of $i$, namely $NN(i)$.

Every particle $i$, described by its position $\vec{r}_i$ and mass $m_i$, is virtually smoothed over a virtual volume of radius $h_i$ using the desired continuous and normalized kernel function $W(\vec{r},h_i)$, usually a continuous function with continuous derivatives over a finite support like a cubic or quintic spline.

The appropriate value of $h_i$ is chosen to satisfy
\begin{equation}
\frac{4}{3} \pi h_i^3 \sumnn m_j W_{ij} = \sumnn m_j
\end{equation}
where we defined $W_ij = W(|\vec{r}_j-\vec{r}_j|,h_i)$, for simplicity.

The first iteration over all particles, is meant to set (and subsequently update) the correct value of $h_i$, as well as computing the value of
\begin{equation}
\rho_i = \sumnn m_j W_{ij}
\end{equation}
representing the density field.

In the second iteration the first derivative
\begin{equation}
\nabla \rho_i = \sumnn m_j \nabla W_{ij} \frac{\rho_j - \rho_i}{\sqrt{\rho_i \rho_j}}
\end{equation}
and the second derivative
\begin{equation}
\nabla^2 \rho_i = \sumnn m_j \nabla^2 W_{ij} \frac{\rho_j - \rho_i}{\sqrt{\rho_i \rho_j}} - \frac{|\nabla \rho_i|^2 }{ \rho_i}
\end{equation}
of the density field are computed, as necessary intermediate quantities.

The third iteration is the one where the typical FDM quantum contribution to acceleration
\begin{equation} 
\label{eq:contribution}
\begin{split}
\nabla Q_i & = \sumnn \nabla Q_{ij} = \\ 
& = \frac{\hslash}{2m_\chi^2} \sumnn \frac{m_j}{\rho_j} \nabla W_{ij} \left( \frac{\nabla^2 \rho_j}{2\rho_j} - \frac{|\nabla \rho_j|^2}{4\rho_j^2} \right)
\end{split}
\end{equation}
is computed.

Even though the equations that constitute the SPH scheme of \FGAS and \AG are same, there are several technical differences in the way these quantities are allocated and computed in practice, among which one is particularly significant and worth mentioning. 

Energy conservation is a requirement that is achievable in many ways in SPH, one of them being ensuring that two-particle contributions to accelerations $\nabla Q_{ij}$ are strictly antisymmetric for $i \longleftrightarrow j$ particle exchange. As stated in \citet{Nori18}, there is no SPH formulation for $\nabla Q_{ij}$ that can be found in order to satisfy this. Nevertheless, differently from \AG, \FGAS solves this problem by splitting equally the two-particle contribution between the two particles, effectively computing $\nabla Q_{i} = \sumnn (\nabla Q_{ij} + \nabla Q_{ji})/2$, thus ensuring energy conservation at the cost of one additional operation. Deviations from energy conservation were not observed in \AG, due to the difficulties of simulating extreme fluctuations in the density field required for them to emerge. Nonetheless, the \FGAS algorithm provide a safer scheme able to prevent such deviations altogether by construction.

\subsection{Simulations}
\label{sec:sims}

To investigate the interplay of FDM and baryonic physics in a variety of contexts, we make use of a set of more than 30 zoom-in simulations. This set is build upon 3 galaxy simulations extracted from the NIHAO database.
NIHAO (Numerical Investigation of Hundred Astrophysical Objects) is one of the largest database of very high resolution simulations \citep{wang2015nihao}, and it has been shown to be very successful in capturing the key processes in galaxy formation and to produce very realistic galaxies \citep[e.g.][]{Dutton2016,Maccio2016, Santos2018, Buck2020}

The NIHAO simulation suite was run using Planck mission cosmological parameters \citep{Planck15}: \( H_0 = 67.1 \, \text{km} \, \text{s}^{-1} \text{Mpc}^{-1} \), the matter, dark energy, radiation, and baryon densities are \( \{ \Omega_m, \Omega_\Lambda, \Omega_r, \Omega_b \} = \{ 0.3175, 0.6824, 0.00008, 0.0490 \} \). The power spectrum normalization and slope are \( \sigma_8 = 0.8344 \) and \( n = 0.9624 \), respectively.
Initial conditions were created (both in CDM and FDM using the GRAFIC2 package \citep{Bertschinger_2001}
and the analysis has been performed using the Amiga Halo Finder (AHF) \citep{2004MNRAS.351..399G, 2009ApJS..182..608K}.
The code \AC \citep{axionCAMB} has been used to compute the initial power spectra in the FDM scenarios and we explore two masses for the FDM models, namely $m_{22} = 2$ and $m_{22} = 8$.

We selected three galaxies from the NIHAO database, namely g6.77e10, g2.63e10 and g9.26e09\footnote{In NIHAO the name of a galaxy represents its total mass in the low resolution N-body simulation and not the one in the high resolution.}, with total masses of 
$9.28\times 10^{10} M_{\odot}$, $2.70 \times 10^{10} M_{\odot}$ and $6.14 \times 10^{9} M_{\odot}$ in the CDM run, and identified in the following with the tags $L$, $M$ and $S$, respectively (see Tab.~\ref{tab:summary}
and the appendix for more information on numerical and space resolution).

This particular mass range was chosen as it represents the turning point that divides --~on a statistical level~-- dark-matter-dominated systems from more massive baryon-dominated ones. In this sense, in system $S$ baryonic contribution is expected to be marginal while in system $L$ baryons play a major role. To further break the degeneracy of the effects of FDM and baryons on galaxy evolution, for every simulation with baryons --~simply referred to as \textit{hydrodynamical} simulations~--, a dark-matter-only (DMO) counterpart is produced.

\section{Results}
\label{sec:results}

In this Section, we present the results obtained from the study of the simulation set presented in the previous Section. We hereby discuss the proprieties of galaxies in the dark and bright sector, detailing the role and impact of each of the specific FDM model as well as the presence and abundance of gas and stars.

\begin{figure*}
\centering
\includegraphics[width=0.91\textwidth]{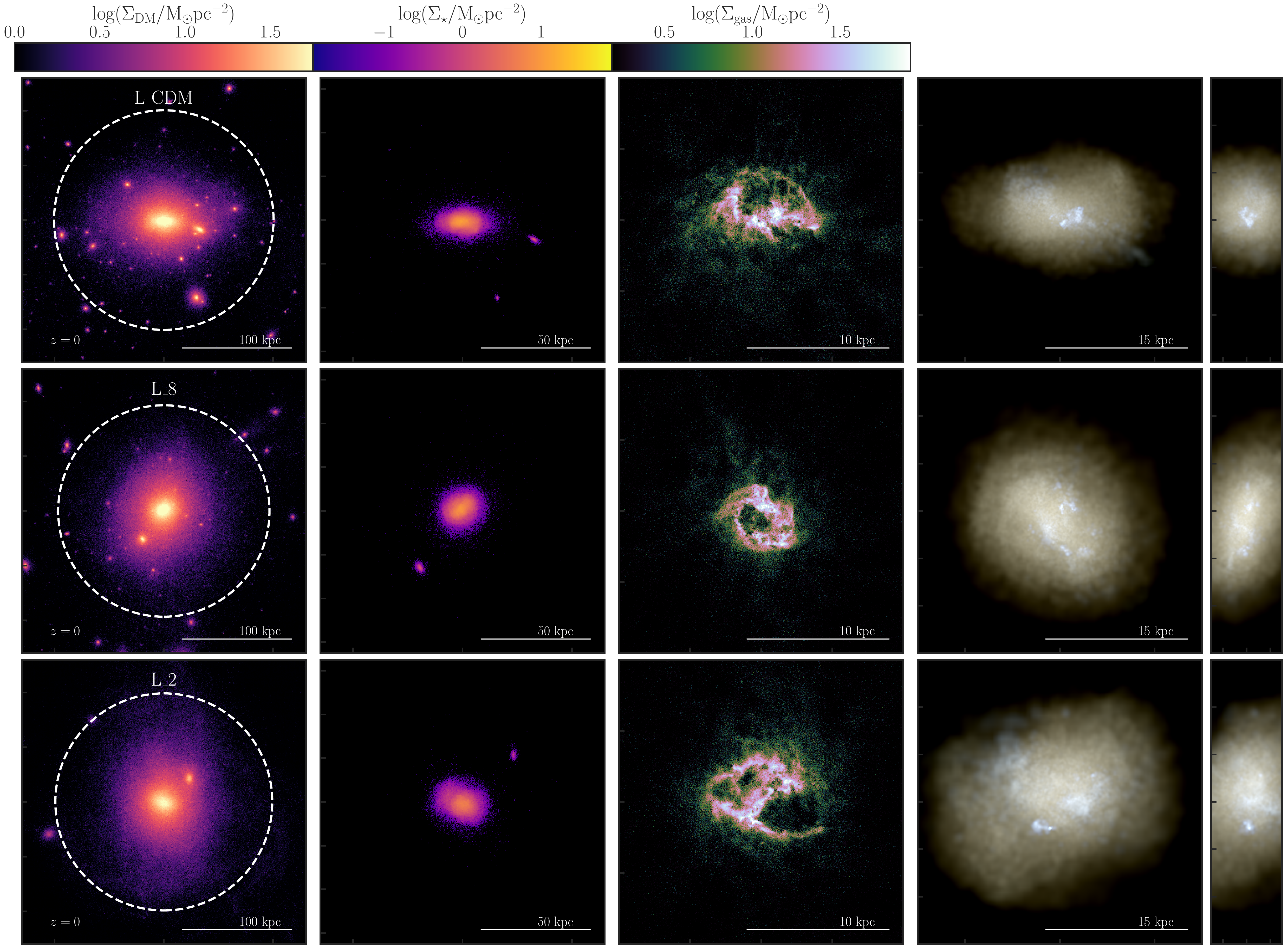}
\includegraphics[width=0.91\textwidth]{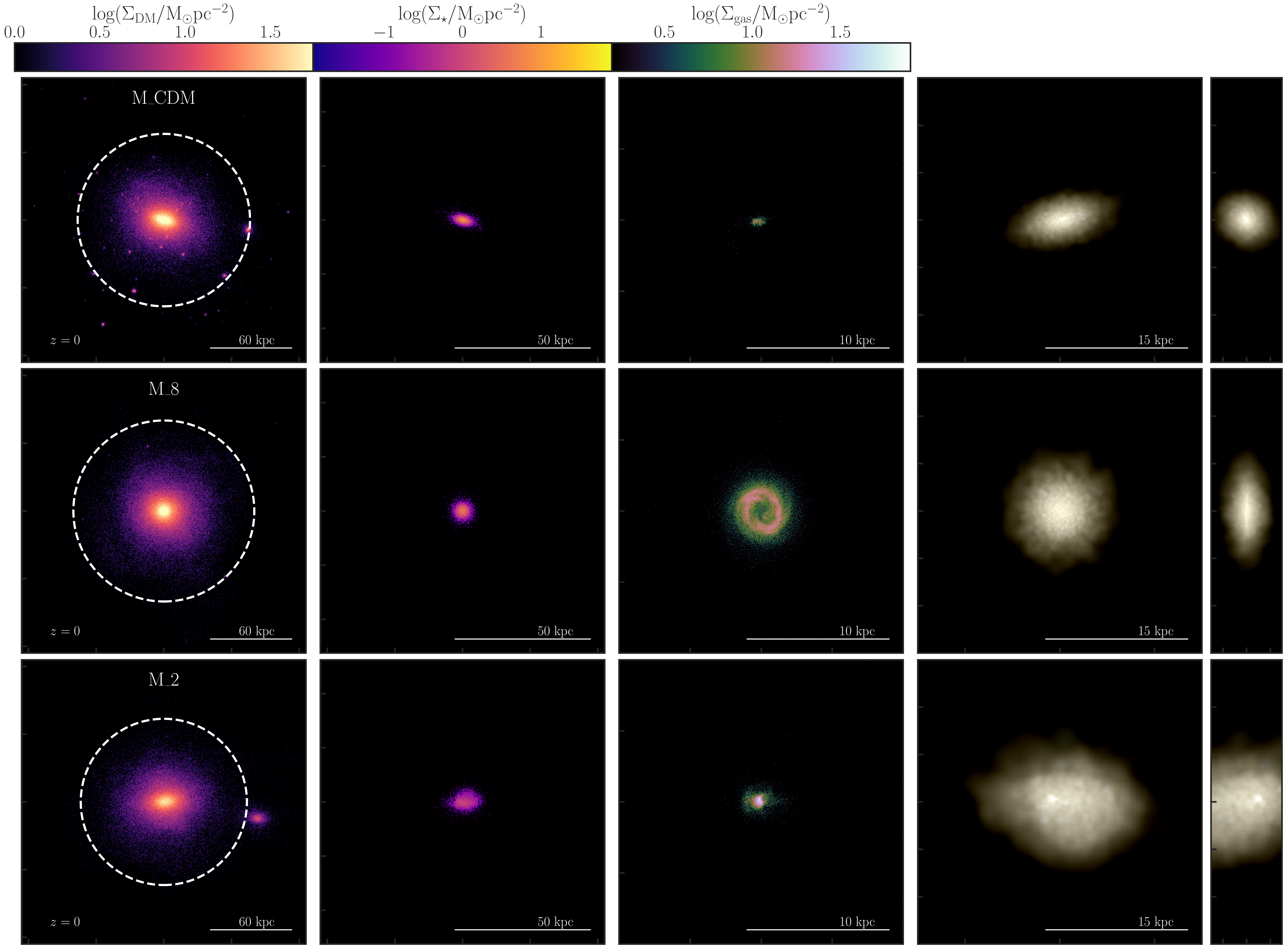}
\end{figure*}

\begin{figure*}
\centering
\includegraphics[width=0.91\textwidth]{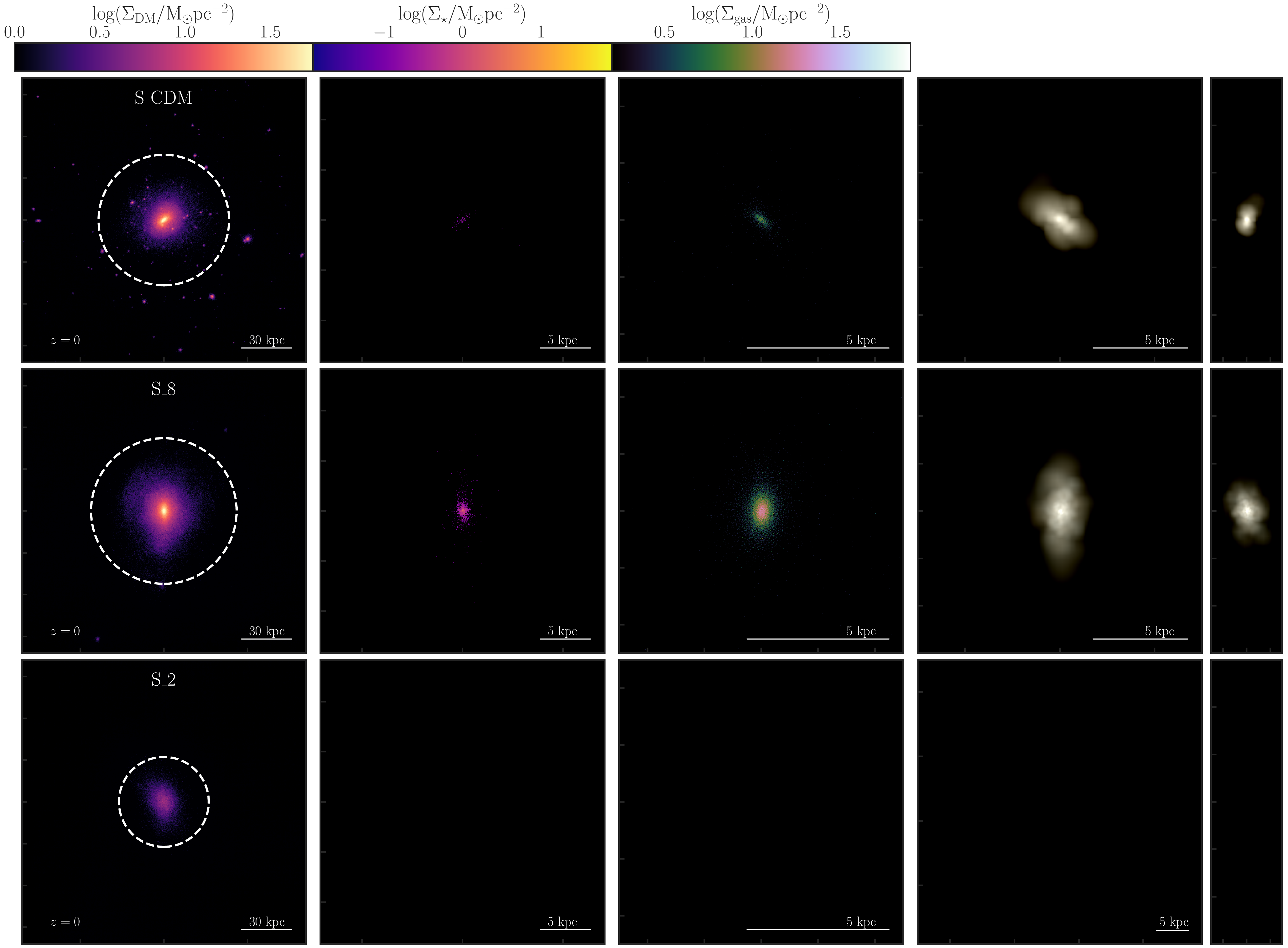}
\caption{Surface density maps viewed face-on as well as stellar face- and edge-on images in the I, V and U wavelength bands for all \FGAS systems (L, M and S from top to bottom). The face-on surface density maps presented column-wise are, from left to right are: dark matter, stars, and gas. The white dashed circle is drawn based on the virial radius.}
\label{fig:maps}
\end{figure*}

A visual impression of the three galaxies in the different runs is presented in figure \ref{fig:maps}, where we show dark matter, stellar and gas maps at different scales, while global properties of the galaxies (like mass and radius) are summarized in Tab.~\ref{tab:summary}.

{As expected the virial mass of the $L$ system decreases (and conversely, the virial radius increases) as the boson mass of the FDM models decrease as well. This reflects the lower concentration of the system induced by an increasingly stronger repulsive force. On the other hand, systems $M$ and --~even more so~-- $S$ feature an increase of the virial mass in the $m_{22} = 8$ models with respect to the CDM case, while for $m_{22} = 2$ the virial mass is lower than its CDM counterpart. Although not immediately intuitive, this is an expected effect of FDM, specifically related to the mass increase of haloes whose virial mass is close to the \textit{threshold} mass $M_t$, representing an estimate of the typical mass below which the number of haloes able to form in FDM statistically deviates from CDM \citep[see][]{Nori19}. Such increase of the virial mass in the $m_{22} = 8$ models is due to the redistribution of mass from the substructures (unable to form due to the FDM interaction) to the main halo structure, which not only compensate but exceeds the redistribution of mass from the central regions to the outskirts of the main structure itself. In fact, it takes a stronger interaction (i.e. a lower value $m_{22} = 2$) for the internal redistribution of matter to take over. In this picture, this increase is not noticeable in system $L$ as some of its subsystems are able to form in all FDM models investigated, thus no such mass transfer effects occur.}
\begin{table*}
\centering
\begin{tabular}{|c|c|c|c|c|c|}
\hline
\textbf{Galaxy} & \textbf{r$_{vir}$ [kpc]} & \textbf{M$_{vir}$ [$10^9$ M$_{\odot}$]}&  \textbf{M$_{dm}$ [$10^9$M$_{\odot}$]}&\textbf{M$_{gas}$ [$10^9$M$_{\odot}$]}& \textbf{M$_{star}$ [$10^7$M$_{\odot}$}]\\
\hline
L\_CDM & 100.4 & 92.8& 85.7& 2.72& 48.4\\
L\_8 & 96.7 & 87.4& 82.0& 2.04& 41.5\\
L\_2 & 99.2 & 84.8& 77.9& 2.92& 31.4\\
\hline
M\_CDM & 63.7 & 27.0& 26.6& 0.04& 4.28\\
M\_8 & 66.8 & 31.0& 30.0& 0.71& 4.14\\
M\_2 & 61.4 & 22.4& 21.9& 0.15& 2.40\\
\hline
S\_CDM & 39.0 & 6.14& 6.03& 0.02& 0.005\\
S\_8 & 43.5 & 8.58& 8.37& 0.07& 0.079\\
S\_2 & 26.9 & 1.96& 1.95& <0.01& 0\\
\hline
\end{tabular}
\caption{Global properties of the simulated systems, further information including mass resolution can be found in the appendix.}
\label{tab:summary}
\end{table*}

\subsection{Dark properties}

\begin{figure*}
\centering
\includegraphics[width=\textwidth,
trim={5.1cm 0.4cm 5.3cm 1.5cm}, clip]{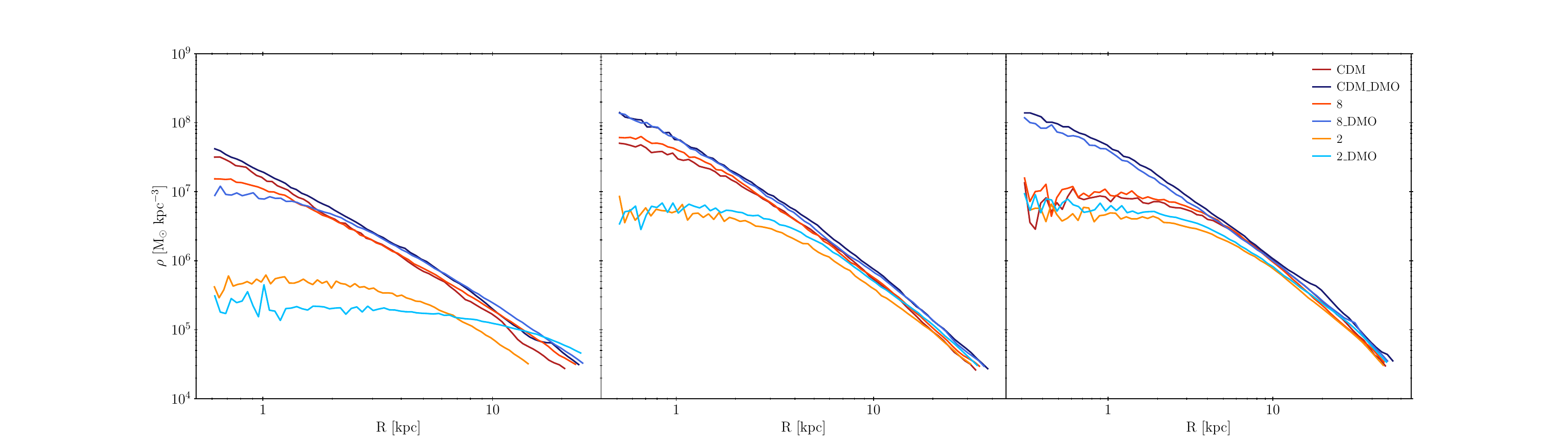}
\caption{Dark matter radial density profiles of all the simulations at $z=0$, gathered by system ($S$, $M$ and $L$ in the left, central and right column, respectively). Dark-matter-only simulations are represented in blue shades, while simulations with baryons with reds.}
\label{fig:dmprof}
\end{figure*}

The first observable presented is the radial dark matter density profile, whose shape is known to be influenced by the presence of baryons in the central region of the halo as well as by a low enough FDM mass particle. In Fig.~\ref{fig:dmprof} we show the density profiles at redshift zero for the  $S$, $M$ and $L$ galaxies, respectively, from left to right, for our three cosmological models (two FDM and one CDM).

It has been shown that, for haloes with a total mass below $10^{12} M_{\odot}$ the sloshing of the gas due to SN feedback can induce a flattening of the otherwise divergent dark matter density profile, thus transforming a so-called \textit{cuspy} profile in a \textit{cored} one \citep[e.g.][]{Tollet15}. The higher the total mass of the system, the higher is --~in general~-- the relative abundance of baryonic-over-dark matter, the stronger the effect. It is possible to observe this in the CDM case: in system $L$ undergoes such baryon-induced core formation process in simulations where baryons are present (i.e. L\_CDM vs L\_CDM\_DMO), unlike dark matter dominated systems like $S$ whose NFW-like profile is unaltered (i.e. S\_CDM vs S\_CDM\_DMO). In system $M$, baryons are enough to change the shape of the profile without eliminating its divergent trend.

Turning to the FDM effects, the typical net repulsive force induces the formation of cores in the dark matter density profile as well. Nevertheless, this effect is pronounced when the strength of the FDM interaction overtakes gravity in the central region of the halo, condition that is generally valid --~for low enough values of the FDM particle mass~-- at the center of less massive systems. These considerations can be easily verified comparing different DMO profiles.

The combined effect of baryons and FDM appears to be not one of addition but rather the resulting dark matter profile is consistent with the most suppressed profile between the one induced by baryons and FDM individually. As an example, let us focus on system $M$: the suppression of the divergence of CDM DMO (M\_CDM\_DMO) profile is light when the sole presence of baryon (M\_CDM) is considered, while the sole FDM suppression effects are dramatic for $m_{22}=2$ (M\_2\_DMO) and barely noticeable for $m_{22}=8$ (M\_8\_DMO). The combination of the two suppression effects results in the final profile to be consistent with the one induced by the individual strongest effect, thus M\_2 and M\_8 profiles being consistent with M\_2\_DMO and M\_CDM ones, respectively.

For the other two systems, the same result is valid. In system $S$, where the presence of baryons is only marginal, the FDM particle mass is the main driver in the formation of the core. Instead, in system $L$, the strong suppression induced by the great abundance of baryons is the main driver for the FDM particle masses considered.

\begin{figure}
%\centering
\includegraphics[width=\columnwidth,trim={0.4cm 0.4cm 0.5cm 0.5cm},clip]{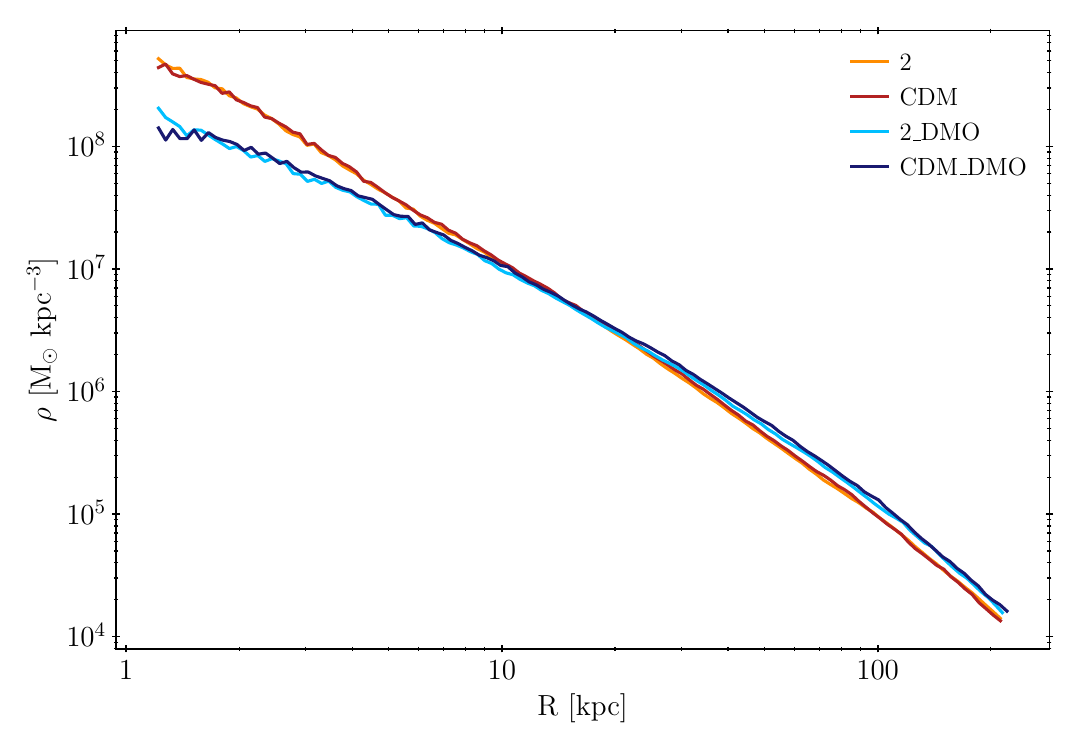}
\caption{Dark matter radial density profiles of an additional system approximately $40$ times more massive than system $L$ at $z=0$.}
\label{fig:dmmassive}
\end{figure}

As a further consistency check, we run three additional simulations based on a more massive system from the NIHAO set (namely g2.79e12) with total mass $3.53\times10^{12} M_{\odot}$, thus approximately $40$ times more massive than system $L$. In line with expectations, in such massive system both baryons and the FDM models investigated are unable to induce the formation of a noticeable core, as it is possible to see in Fig.~\ref{fig:dmmassive}. As this system does not feature any relevant deviation between FDM and CDM, it will not be included in the analysis presented in the next sections. 

\begin{figure*}
\centering
\includegraphics[width=0.517\textwidth,trim={0.4cm 0.4cm 0.5cm 0.5cm},clip]{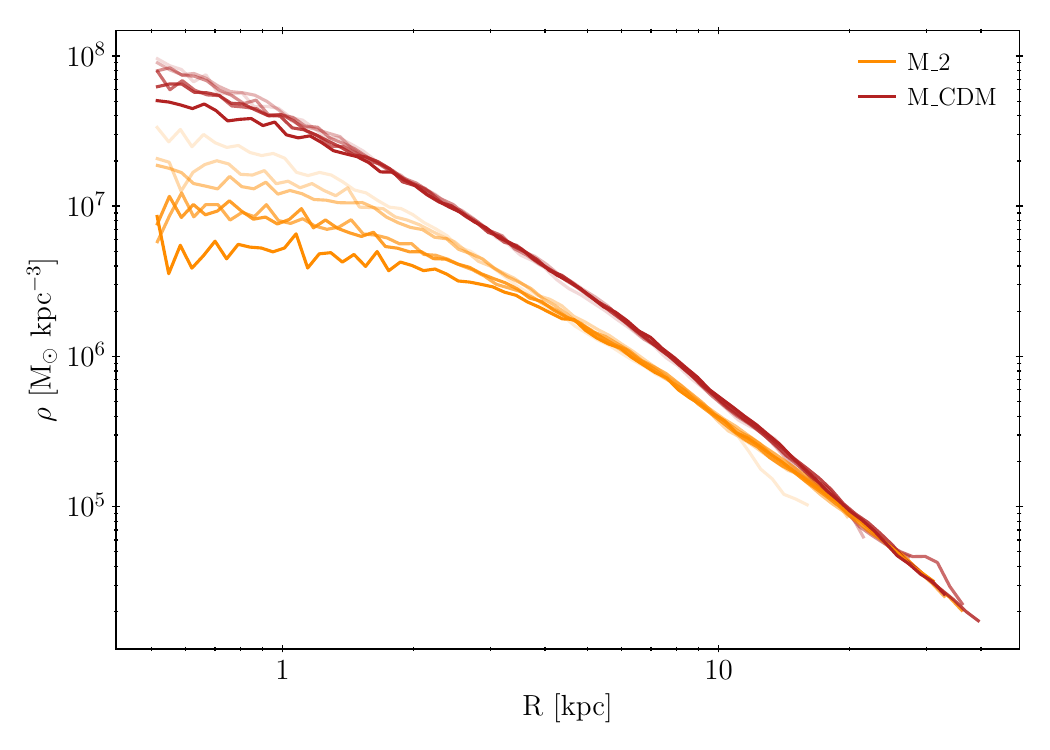}
\includegraphics[width=0.477\textwidth,trim={0.4cm 0.4cm 0.5cm 0.5cm},clip]{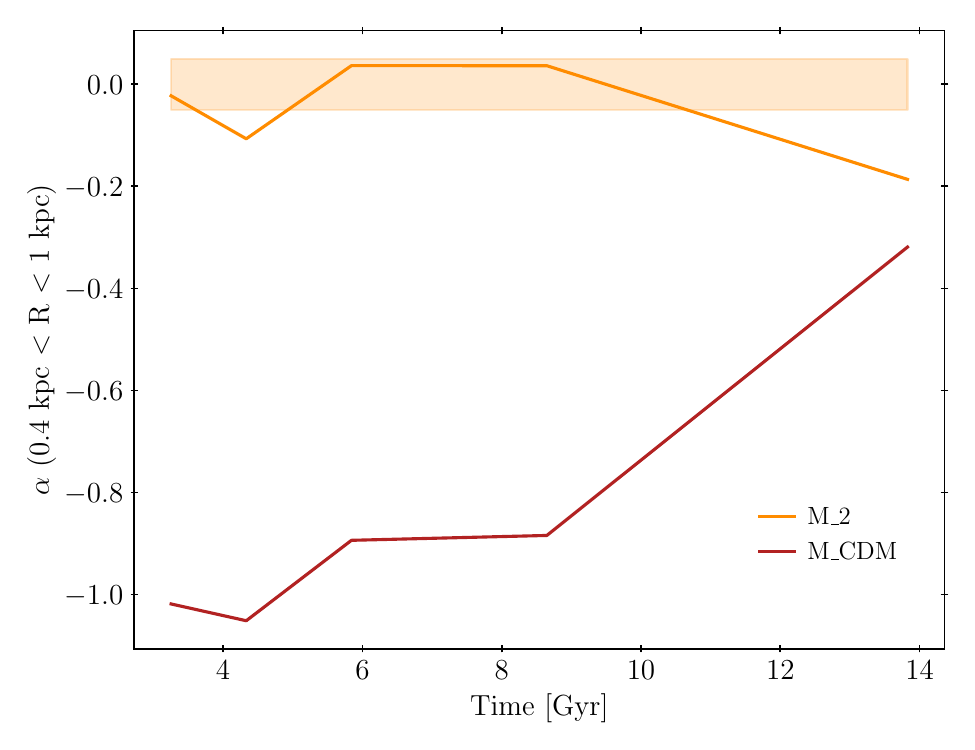}
\caption{Evolution in redshift of the dark matter radial density profiles of system $M$, for the two specific models M\_CDM and M\_2. Redshifts displayed are $z=\{0, 0.5, 1, 1.5, 2, 2.5\}$ in lighter shades from high to low redshift. In the right hand plot the value of the slope $\alpha$ in FDM, is only shown for negative values, since positive dynamically unstable values are due to noise in the profile (the shaded band represents this simplification).}
\label{fig:dmevo}
\end{figure*}

Although FDM and baryons both are able to produce a core at low redshift, this is not the case at higher redshifts. In fact, the core formation process induced by baryons needs time for gas and stars to form and accumulate in the center and alter dark matter distribution, thus it is a typical low redshift feature. On the contrary, FDM net repulsive force is stronger at higher redshift and gets weaker as time goes by, thus prompting core formation well before baryons would. To appreciate this, dark matter density profiles from redshift $z=2.5$ to $z=0.0$ for the $M$ system are gathered in the left panel of Fig.~\ref{fig:dmevo}, where a core is present at higher redshift only in the FDM $m_{22} = 2$ model but not in the CDM one. To quantitatively represent the redshift evolution of the core feature, the evolution of the profile slope $\alpha$ calculated between $0.4$kpc and $1$kpc is reported in the right panel of Fig.~\ref{fig:dmevo}. It is very clear that while in CDM the slope $\alpha$ slowly grows from negative values to zero \citep[see also][]{Tollet15}, as baryons induce the formation of a core over time, values of the slope in FDM fluctuate around zero, consistently with a core at all the redshifts considered. 

These evidences confirm that an early onset of a core in the dark matter profile can only be linked to FDM and not to baryons. Thus, indirect observations of cores in the dark matter profile at higher redshift could be a valuable probe of the FDM paradigm validity.

\begin{figure*}
\centering
    
\includegraphics[width=\textwidth,
trim={5.4cm 0.4cm 5.3cm 1.8cm}, clip]{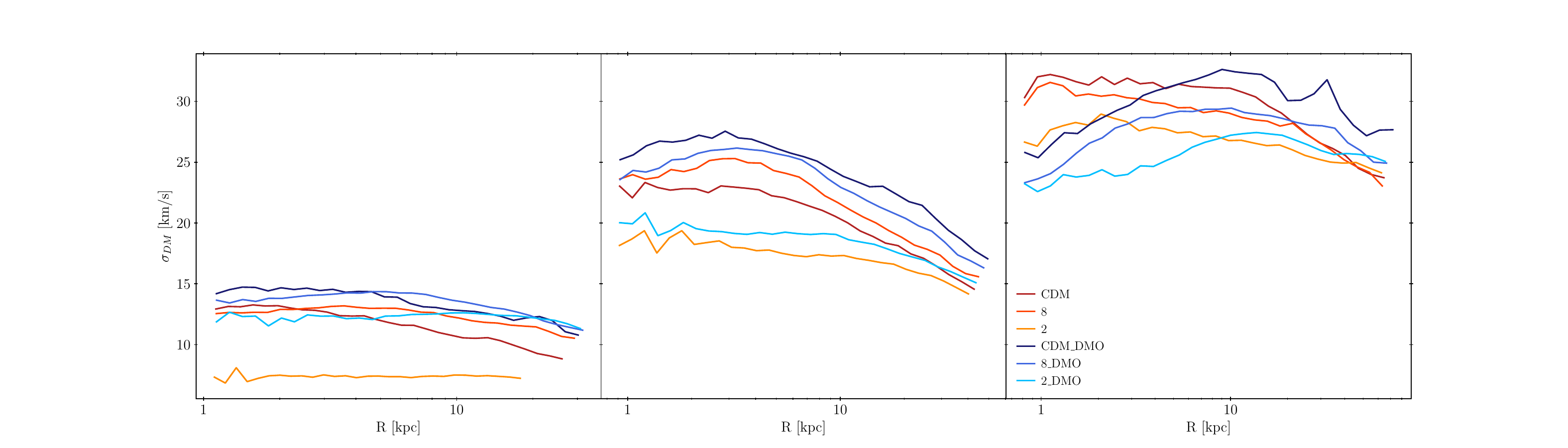}
\caption{Dark matter velocity dispersion profiles of all the simulations at $z=0$, gathered by system ($S$, $M$ and $L$ in the left, central and right column, respectively). Dark-matter-only simulations are represented in blue shades, while simulations with baryons with reds.}
\label{fig:vdprof}
\end{figure*}

In Fig.~\ref{fig:vdprof}, the velocity dispersion radial profiles is presented, where the plot configuration and color scheme are the same as in Fig.~\ref{fig:dmprof}. The impact of FDM and baryons on the velocity dispersion profile has a similar flattening and lowering effects in terms of shape and absolute value respectively, although being caused by entirely different physical reasons. The former is the reflection of the overall balance of attractive and repulsive forces --~i.e. sourced by the gravitational and quantum potential, respectively~-- whereas the latter emerges from the altered central dark matter density distribution induced by baryons. The combination of these two effects seems to add up, as the hydrodynamical counterpart is flatter in the center than its DMO counterpart, for all model investigated. {It is worth to notice that, following the higher values of $M_{vir}$ in system $M$ and $S$ for the $m_{22}=8$ FDM model, the velocity dispersion in these cases have statistically higher absolute values than their CDM counterparts, as this observable is closely linked to virial mass.}

\subsection{Visible properties}

\begin{figure*}
\centering
\includegraphics[width=\textwidth,
trim={5.1cm 0.4cm 5.3cm 1.5cm}, clip] {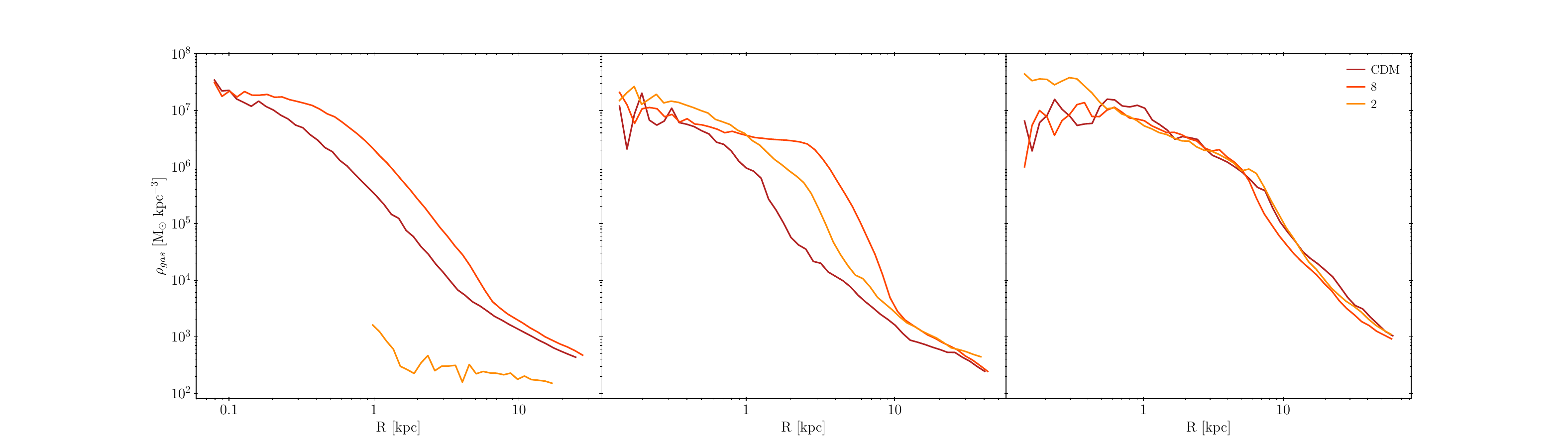}
\includegraphics[width=\textwidth,
trim={5.1cm 0.4cm 5.3cm 1.5cm}, clip] {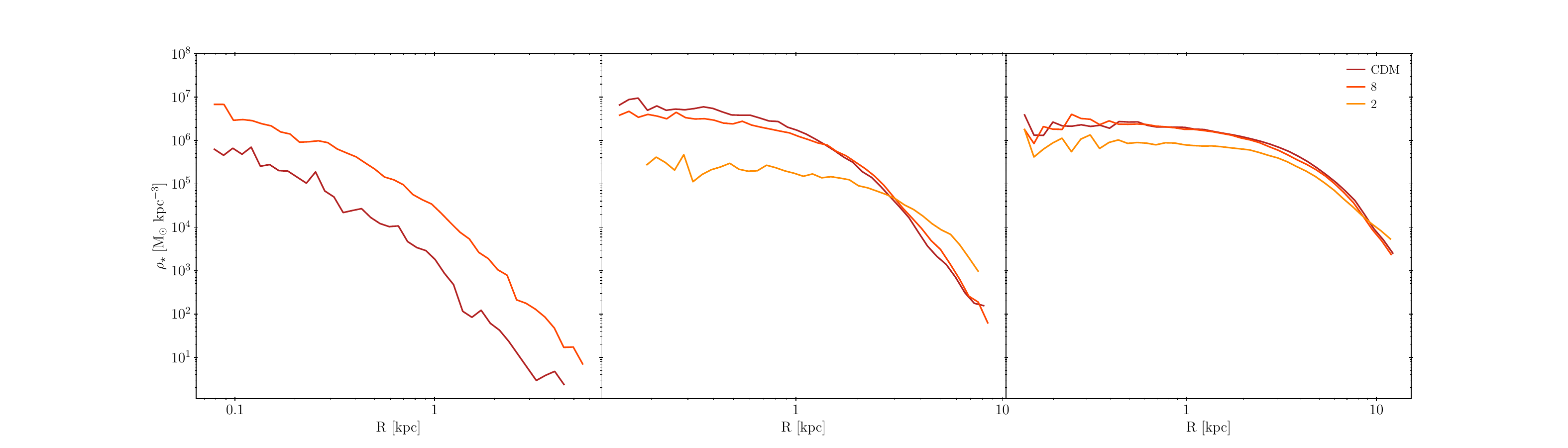}
\caption{Gas (first row) and star (second row) density profiles, gathered by columns for the $S$, $M$ and $L$ systems, respectively.}
\label{fig:gsprof}
\end{figure*}

Turning to baryonic observables, let us discuss the radial density profile of gas and stars, gathered in the top and bottom row of Fig.~\ref{fig:gsprof}, respectively.

The gas radial profiles in system $L$ are substantially consistent across the CDM and the two FDM models. The star profiles, already cored in shape in the CDM models, exhibit a substantial deviation from CDM only in the $m_{22}=2$ FDM model, with a cored profile with the same shape but a lower central density of approximately $\sim0.5$ factor.

In system $M$, although the gas profiles differ in shape in the across models, a deviation in the star profiles is observed only in $m_{22}=2$ FDM model, with a more extended cored profile at a much lower central value (of approximately a $\sim0.1$ factor). Thus, in both $L$ and $M$ systems, the star profile in the $m_{22}=8$ model is fully consistent with CDM.

Instead, in system $S$, gas and star profiles vary significantly between models. On one hand, in the $m_{22}=2$ FDM model, gas is barely present and the star component lies below resolution, virtually exhibiting a total lack of stars. On the other hand, both gas and star content exceeds the CDM one in the $m_{22}=8$ model, as a reflection of the higher overall mass redistribution from substructures to the central region discussed at the beginning of the Section.

\begin{figure*}
\centering
\includegraphics[width=\textwidth,
trim={4.9cm 0.4cm 5.3cm 1.7cm}, clip]{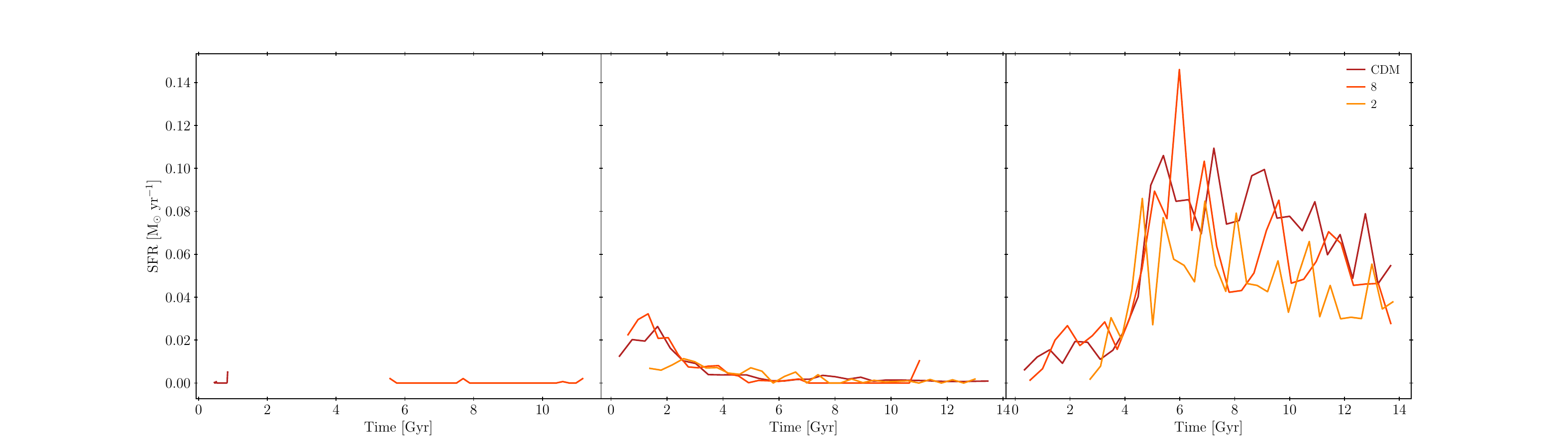}
\caption{Star formation histories for the $S$, $M$ and $L$ systems, respectively.}
\label{fig:sfh}
\end{figure*}

{To better understand the similarities and variations in the star distribution between models at redshift $z=0$, it is valuable to investigate the rate of star formation in time by studying the Star Formation Histories (SFHs), presented in Fig.~\ref{fig:sfh}.}

{In system $M$ and $L$, where a relevant variation in the star distribution is observed only in the FDM $m_{22}=2$ model, FDM is able to alter the star formation history by delaying the initial moment and overall production of stars onward. This is consistent with FDM }
the suppressed power spectrum in FDM models, that delays structure formation in a way similar to what has been observed in simulations based on warm dark matter \citep[see for example][]{Schneider_etal_2012, Stoychev2019, Maccio2019}.

The situation is somehow reversed in the $S$ system, that has  more stars and a more prolonged SFR in the 
FDM $m_{22}=8$ case with respect to CDM. In fact while the CDM run did start to produce stars earlier, then the subsequent feedback related to this early burst of SF was enough to expel the majority of the gas and prevent any further star formation. While in the $m_{22}=8$ run SF did not happen till $z \approx 1$, and by that time the halo was massive enough to successfully "survive" the first SN explosions, and retain enough gas to have a non zero SFR for about 5 Gyrs, ending up with more stars than its CDM counterpart. 
Conversely in the $m_{22}=2$ case, structure formation was so delayed that the halo never reached the critical  mass for SF and did not form any stars \citep{Benitez-Llambay2020}.

Our results show that FDM has generally a negative impact  on star formation, mostly by delaying its onset and reducing its overall value, but it can also act as a positive feedback in some limited cases, especially for systems at the very low masses that have a very stochastic SF histories \citep{Frings2017MNRAS}, confirming the complex picture of cascading effects that characterize FDM scenarios. This poses the question on whether early constraints on FDM that used semi-analytic extrapolations to derive stellar distributions in dwarf galaxies should be reconsidered.

\section{Conclusions}
\label{sec:conclusions}

{
In this work, we presented the N-body code \FGAS, able to correctly simulate evolving astrophysical systems in the Fuzzy Dark Matter (FDM) scenario. The specific Smoothed Particle Hydrodynamic routines that computes the additional force sourced by the typical Quantum Potential of FDM dynamics, a refined version of the ones found in \AG, are in \FGAS complemented with a state of the art ensemble of sophisticated routines related to cooling and star formation, black hole physics, supernova feedback. This makes \FGAS the first code able to properly simulate complex galactic systems in the FDM scenario with a full hydrodynamic description.
}

{Results from the first simulation set presented in this work showed that the impact of FDM dynamics and baryonic physics on the typical observables that describe galactic systems are intrinsically coupled. However, by comparing different simulation setups that exclude/include both effects singularly, it has been possible to single out specific contributions and effects.
}

{In particular, we verified that the presence of baryons and a low-enough boson FDM mass are both able to form a core in the dark matter radial profile. However, the two are able to induce the formation of a core in different contexts: FDM dynamics form cores in all haloes, which are typically more pronounced for low-mass systems and at high redshifts; baryons more effective in inducing a core formation, especially at low-redshift, in systems within a specific range of mass total of [$10^{10}-10^{11} M_\odot$] \citep{2020MNRAS.495L..46M}. Whenever this two contexts overlap, the final effect is consistent with the strongest or with a combination of the two, depending on the observable.}

{
When detailing the deviation from standard CDM scenario, FDM exhibits strong effects in the dark sector, as expected. However, it is extremely interesting to notice that despite this the observable related to stars are only partially affected, when considering reasonable values of boson FDM mass. This suggests that from an observational point of view, a rather important FDM variation in the dark sector can lead to similar results to CDM when restricting to properties directly or indirectly measured from stellar observables, especially at low redshift.
}

In other words, our results suggest that baryons are affected by the changes induced by FDM in a rather inefficient way, thus making FDM scenarios a viable model to describe alteration or tensions in the dark sector while not disrupting the consistency of the results obtained in the visible sector. This is a first pilot study and we plan to expand our simulations portfolio in FDM both in mass (following the NIHAO approach) and both in redshift, extending FDM to our new large simulations suite HELLO \citep{Waterval2024}.

\section*{Acknowledgments}

This material is based upon work supported by Tamkeen under the NYU Abu Dhabi Research Institute grant CASS.
The authors gratefully acknowledge the High Performance Computing resources at New York University Abu Dhabi, where all simulations have been performed. The authors warmly thank Dr. Stefan Waterval for the insightful discussions, his coding guidance and the support in producing the images in Fig.~\ref{fig:maps}.

\section*{Data Availability}
The data underlying this article will be shared on reasonable request to the corresponding author.

%%%%%%%%%%%%%%%%%%%%%%%%%%%%%%%%%%%%%%%%%%%%%%%%%%



\bibliographystyle{mnras}
\bibliography{BIB,baldi_bibliography} % if your bibtex file is called example.bib

%%%%%%%%%%%%%%%%%%%%%%%%%%%%%%%%%%%%%%%%%%%%%%%%%%

%%%%%%%%%%%%%%%%% APPENDICES %%%%%%%%%%%%%%%%%%%%%

\appendix
\section{Galaxy Parameters}

\begin{table*}
    \centering
    \begin{tabular}{|c|c|c|}
        \hline
        \textbf{Galaxy} & \textbf{$\epsilon_{star}$ [kpc]} & \textbf{$\epsilon_{dm}$ [kpc]} \\
        \hline
        L\_CDM, L\_8, L\_2 & 0.1326 & 0.3105 \\
        %L\_8 & 0.1326 & 0.3105 \\
        %L\_2 & 0.1326 & 0.3105 \\
        M\_CDM, M\_8, M\_2 & 0.1326 & 0.3105 \\
        %M\_8 & 0.1326 & 0.3105 \\
        %M\_2 & 0.1326 & 0.3105 \\
        S\_CDM, S\_8, S\_2 & 0.07461 & 0.1746 \\
        %S\_8 & 0.07461 & 0.1746 \\
        %S\_2 & 0.07461  & 0.1746 \\
        \hline
    \end{tabular}
    \caption{\emph{\FGAS} Galaxy softening lengths for the star and DM particles.}
    \label{tab:galaxy_softening_properties}
\end{table*}

\begin{table*}
    \centering
    \begin{tabular}{|c|c|c|c|c|c|c|}
        \hline
        \textbf{Galaxy} & \textbf{M$_{halo}$ [$10^9$ M$_{\odot}$]} & \textbf{M$_{dm}$ [$10^9$ M$_{\odot}$]} & \textbf{SFR [M$_{\odot}$ 100 Myr$^{-1}$]} & \textbf{n$_{stars}$} & \textbf{n$_{dm}$} & \textbf{n$_{gas}$} \\
        \hline
        L\_CDM & 92.8 & 85.7 & 0.0776 & 204017 & 1333784 & 580301 \\
        L\_8 & 87.4 & 82.0 & 0.0386 & 175939 & 1275307 & 439736 \\
        L\_2 & 84.8 & 77.9 & 0.0838 & 131505 & 1211598 & 574337 \\
        M\_CDM & 27.0 & 26.6 & 0 & 18388 & 414291 & 26044 \\
        M\_8 & 31.0 & 30.0 & 0 & 17825 & 467263 & 83178 \\
        M\_2 & 22.4 & 21.9 & 0 & 10335 & 340075 & 43852 \\
        S\_CDM & 6.14 & 6.03 & 0 & 132 & 527275 & 53954 \\
        S\_8 & 8.58 & 8.37 & 0 & 1908 & 731472 & 103235 \\
        S\_2 & 1.96 & 1.95 & 0 & 0 & 170681 & 5230 \\
        \hline
    \end{tabular}
    \caption{\emph{\FGAS} galaxy properties.}
    \label{tab:galaxy_properties}
\end{table*}

\begin{table*}
    \centering
    \begin{tabular}{|c|c|c|c|c|c|c|}
        \hline
        \textbf{Galaxy} & \textbf{r$_{vir}$ [kpc]} & \textbf{M$_{vir}$ [$10^9$ M$_{\odot}$]} & \textbf{n$_{vir}^{total}$} & \textbf{n$_{vir}^{stars}$} & \textbf{n$_{vir}^{dm}$} & \textbf{n$_{vir}^{gas}$} \\
        \hline
        L\_CDM & 100.4 & 92.8 & 2118102 & 204017 & 1333784 & 580301 \\
        L\_8 & 96.7 & 87.4 & 1890982 & 175939 & 1275307 & 439736 \\
        L\_2 & 99.2 & 84.8 & 1917440 & 131505 & 1211598 & 574337 \\
        M\_CDM & 63.7 & 27.0 & 458723 & 18388 & 414291 & 26044 \\
        M\_8 & 66.8 & 31.0 & 568266 & 17825 & 467263 & 83178 \\
        M\_2 & 61.4 & 22.4 & 394262 & 10335 & 340075 & 43852 \\
        S\_CDM & 39.0 & 6.14 & 581361 & 132 & 527275 & 53954 \\
        S\_8 & 43.5 & 8.58 & 836615 & 1908 & 731472 & 103235 \\
        S\_2 & 26.9 & 1.96 & 175911 & 0 & 170681 & 5230 \\
        \hline
    \end{tabular}
    \caption{\emph{\FGAS} galaxy virial properties.}
    \label{tab:galaxy_virial_properties}
\end{table*}

\begin{table*}
    \centering
    \begin{tabular}{|c|c|c|c|c|}
        \hline
        \textbf{Galaxy} & \textbf{M$_{star}$ [$10^7$ M$_{\odot}$]} & \textbf{M$_{tot}$ [$10^9$ M$_{\odot}$]} & \textbf{M$_{star}$/M$_{tot}$} \\
        \hline
        L\_CDM & 48.4 & 24.6 & 1.96E-02 \\
        L\_8 & 41.5 & 21.9 & 1.89E-02 \\
        L\_2 & 31.4 & 18.6 & 1.69E-02 \\
        M\_CDM & 4.28 & 8.84 & 4.85E-03 \\
        M\_8 & 4.14 & 10.0 & 4.14E-03 \\
        M\_2 & 2.40 & 4.99 & 4.80E-03 \\
        S\_CDM & 0.00537 & 1.65 & 3.25E-05 \\
        S\_8 & 0.0793 & 1.91 & 4.16E-04 \\
        S\_2 & 0 & 0.202 & 0 \\
        \hline
    \end{tabular}
    \caption{\emph{\FGAS} Galaxy stellar and total mass properties, computed within 20\% of the virial radius of the galaxy.}
    \label{tab:galaxy_stellar_mass_properties}
\end{table*}

\begin{table*}
    \centering
    \begin{tabular}{|c|c|c|c|c|c|c|}
        \hline
        \textbf{Galaxy} & \textbf{M$_{gas}$ [$10^9$ M$_{\odot}$]} & \textbf{M$^{gas}_{cold}$ [$10^7$ M$_{\odot}$]} & \textbf{M$^{gas}_{cool}$ [$10^7$ M$_{\odot}$]} & \textbf{M$^{gas}_{warm}$ [$10^6$ M$_{\odot}$]} & \textbf{M$^{gas}_{hot}$ [$10^6$ M$_{\odot}$]} \\
        \hline
        L\_CDM & 2.72 & 143 & 123 & 55.9 & 1.85 \\
        L\_8 & 2.04 & 46.2 & 154 & 41.3 & 1.75 \\
        L\_2 & 2.92 & 53.1 & 233 & 52.9 & 0.0453 \\
        M\_CDM & 0.0411 & 1.36 & 2.66 & 0.94 & 0 \\
        M\_8 & 0.706 & 41.0 & 28.9 & 6.49 & 0 \\
        M\_2 & 0.154 & 4.96 & 10.4 & 0.32 & 0 \\
        S\_CDM & 0.0161 & 1.74 & 1.44 & 0.00 & 0 \\
        S\_8 & 0.071 & 19.3 & 5.17 & 0.0024 & 0 \\
        S\_2 & <0.01 & 0 & 0.0167 & 0 & 0 \\
        \hline
    \end{tabular}

    \caption{\emph{\FGAS} Galaxy gas properties, computed within 20\% of the virial radius of the galaxy. Cold: T < $10^4$ K; Cool: $10^4$ K < T < $10^5$ K; Warm: $10^5$ K < T < 5 x $10^6$ K; Hot: T > $10^6$ K.}
    \label{tab:galaxy_gas_properties}
\end{table*}

%%%%%%%%%%%%%%%%%%%%%%%%%%%%%%%%%%%%%%%%%%%%%%%%%%

% Don't change these lines
%\bsp	% typesetting comment
\label{lastpage}
\end{document}